\documentclass[showpacs,preprintnumbers,amsmath,amssymb,prd]{revtex4}
\usepackage{dcolumn}
\usepackage{bm}
\newcommand{\be}{\begin{equation}}
\newcommand{\ee}{\end{equation}}
\newcommand{\ba}{\begin{eqnarray}}
\newcommand{\ea}{\end{eqnarray}}
\newcommand{\nn}{\nonumber}
\def\etap{{\eta^{(\prime)}}}

\begin{document}

\title{Branching ratios and CP asymmetries in charmless nonleptonic $B$ decays to radially 
excited mesons}

\author{G. Calder\'on}
\email{german.calderon@uadec.edu.mx}
\affiliation{Facultad de Ingenier\'ia Mec\'anica y El\'ectrica,
Universidad Aut\'onoma de Coahuila, C.P. 27000, Torre\'on, Coahuila, M\'exico}

\begin{abstract}
Nonleptonic two body $B$ decays including radially excited $\pi(1300)$ or $\rho(1450)$ mesons 
in the final state are studied using the framework of generalized naive factorization approach. 
Branching ratios and CP asymmetries of $B\to P\pi(1300)$, $B\to V\pi(1300)$, $B\to P\rho(1450)$ 
and $B\to V\rho(1450)$ decays are calculated, where $P$ and $V$ stand for pseudoscalar and vector 
charmless mesons. Form factors for $B\to \pi(1300)$ and $B\to \rho(1450)$ transitions 
are estimated in the improved version of the Isgur-Scora-Grinstein-Wise quark model. 
In some processes, CP asymmetries of more than $10\%$ and branching ratios of $10^{-5}$ 
order are found, which could be reached in experiments.
\end{abstract}

\pacs{13.25.Hw, 12.39.Jh, 11.30.Er}

\maketitle

\section{Introduction}

Most of the research on nonleptonic two body $B$ decays have concentrated in processes where the 
final mesons are ground states or angular orbital excitations \cite{babarbook}. Radial excitations 
can be produced in $B$ decays. Decays involving radially excited mesons could be an alternative 
to those more traditionally studied, given additional and complementary information.

The physics involved in nonleptonic two body $B$ decays allows to study the interplay 
of QCD and electroweak interactions, to search for CP violation, and over constrain
Cabbibo-Kobayashi-Maskawa (CKM) parameters in a precision test of the standard model 
\cite{babarbook, bigibook}.

In the quark model, mesons are $q\bar q'$ bound states of quark $q$ and antiquark 
$\bar q'$. The $q\bar q'$ state has an orbital angular momentum $l$ and spin $J$.
Pseudoscalar and vector mesons have orbital angular momentum $l=0$. The angular orbital 
excitations: scalar, axial vector and tensor mesons have $l=1$. Mesons can be classified 
in spectroscopy notation by $n^{2s+1}l_J$, where $s=0$ or $1$ for parallel or antiparallel 
quarks $q$ and $\bar q^\prime$, respectively. Radial excitations are denoted by the principal
quantum number $n$. 

The radial excitations $\pi(1300)$ and $\rho(1450)$, with principal quantum number
$n=2$, $u$ and $d$ quark content, can be produced in nonleptonic two body $B$ decays.
In spectroscopy notation $\pi(1300)$ is denoted by $2^1S_0$ and $\rho(1450)$ 
by $2^3S_1$. To simplify, we denote $\pi(1300)$ by $\pi'$ and $\rho(1450)$ by $\rho'$. 

In Ref. \cite{diehl}, the authors interested in factorization breaking effects in $B$ decays,
consider $B$ decays to final states with small decay constants, such as $\bar B^0\to D^+\pi'^-$  
and $\bar B^0\to D^{*+}\pi'^-$ decays. 

Production of charmless radially excited vector mesons in nonleptonic two body $B$ decays is 
considered in Ref. \cite{lipkin02}. The authors make a prediction for the ratio 
$Br(B\to \rho' \pi)/Br(B\to \rho\pi)$. This ratio is given in terms of the form factor 
$A_0$, which is calculated in a constituent quark model \cite{lipkin02}. We compare our 
calculations with their result and experimental data available.

In this paper, we present a study on the exclusive modes $B\to P\pi'$, $B\to V\pi'$, 
$B\to P\rho'$ and $B\to V\rho'$, where $P$ and $V$ are the pseudoscalar and vector mesons, 
$\pi$, $\eta$, $\eta'$ and $K$ and $\rho$, $\omega$, $K^*$ and $\phi$, respectively. We compute 
branching ratios of these processes using the effective weak Hamiltonian, with tree and penguin 
contributions. Matrix elements are calculated in the generalized naive factorization 
approach \cite{ali99, ali}. The form factors for $B\to P$ and $B\to V$ transitions are calculated 
in the Bauer-Stech-Wirbel (WSB) model \cite{wsb} and Light-Cone-Sum-Rule (LCSR) approach \cite{ball}. 
Form factors for $B\to \pi'$ and $B\to \rho'$ transitions are calculated in the improved version of 
the Isgur-Scora-Grinstein-Wise (ISGW) quark model, called ISGW2 model \cite{isgw89,isgw95}.

We also calculate CP violating asymmetries in the framework of generalized naive factorization 
approach \cite{aliCP}. CP asymmetries allow to determine interior angles of the unitary triangle 
and test the unitarity of the CKM matrix. Specifically, in this work, we calculate direct CP violation 
for charged $B^\pm$ decays and CP asymmetries for neutral $B^0(\bar B^0)$ decays. For some channels 
asymmetries of order $10 \%$ are found.

The decay constants $f_{\pi'}$ and $f_{\rho'}$ are not well determined input parameters.
The range of values for the $f_{\pi'}$ decay constant obtained from different methods and 
its impact in channels $B\to P\pi'$ and $B\to V\pi'$ are discussed in this work. Some 
branching ratios are sensitive to the decay constants $f_{\pi'}$ and $f_{\rho'}$. This fact will 
allow to determine decay constants by experiment in cases where branching ratios are measured.

This paper is organized as follows. In Sec. II, we present the framework used to calculate 
branching ratios of nonleptonic two body $B$ decays, effective Hamiltonian and the generalized
naive factorization approach. Input parameters, mixing schemes, decay constants, and form factors 
are discussed in Sec. III. In Sec. IV, we discuss the amplitudes involving radially excited 
mesons and calculate numerical results for branching ratios. CP violating asymmetries for charged 
and neutral channels are presented in Sec. V. Our conclusions are given in Sec. VI. In the 
appendices, we give the amplitudes for $B\to P\pi'$, $V\pi'$,  $P\rho'$ and $V\rho'$ processes.

\section{Framework}

\subsection{Effective Hamiltonian}

The framework to study $B$ decays is the effective weak Hamiltonian \cite{buras96}.
For $\Delta B=1$ transitions, it is written as

\ba H_{eff} &=& \frac{G_F}{\sqrt{2}}\Bigg[\sum_{j=u,c} V_{jb}V^*_{jq}
\Big(C_1(\mu) O^j_1(\mu) + C_2(\mu) O^j_2(\mu)\Big)
- V_{tb}V^*_{tq}\Bigg(\sum^{10}_{i=3}C_i(\mu) O_i(\mu)
\Bigg) \Bigg] + H.c.,\ea

\noindent where $G_F$ is the Fermi constant, $C_i(\mu)$ are Wilson coefficients at the 
renormalization scale $\mu$, $O_i(\mu)$ are local operators and $V_{ij}$ are the respective
CKM matrix elements involved in the transitions. The local operators for $b\to q$ transitions 
are

\ba O^j_1 &=& \bar q_\alpha\gamma^\mu L u_\alpha \cdot
\bar u_\beta \gamma_\mu L b_\beta \nn\\
O^j_2 &=& \bar q_\alpha\gamma^\mu L u_\beta \cdot
\bar u_\beta \gamma_\mu L b_\alpha \nn\\
O_{3(5)} &=& \bar q_\alpha\gamma^\mu L b_\alpha \cdot
\sum_{q'} \bar q'_\beta \gamma_\mu L(R) q'_\beta \nn\\
O_{4(6)} &=& \bar q_\alpha\gamma^\mu L b_\beta \cdot
\sum_{q'} \bar q'_\beta \gamma_\mu L(R) q'_\alpha \nn\\
O_{7(9)} &=& \frac{3}{2}\bar q_\alpha\gamma^\mu L b_\alpha \cdot
\sum_{q'} e_{q'}\bar q'_\beta \gamma_\mu R(L) q'_\beta \nn\\
O_{8(10)} &=& \frac{3}{2}\bar q_\alpha\gamma^\mu L b_\beta \cdot
\sum_{q'} e_{q'}\bar q'_\beta \gamma_\mu R(L) q'_\alpha , \ea

\noindent where $q=d$ or $s$, $O^j_1$ and $O^j_2$ are the current-current operators ($j=u,c$), 
$O_3 - Q_6$ the QCD penguins operators and $Q_7 - Q_{10}$ the electroweak 
penguins operators. The indexes $\alpha$ and $\beta$ mean $SU(3)$ color degrees,   
$L$ and $R$ are the left and right projector operators, respectively.
The sum extends over active quarks $u$, $d$, $s$ and $c$ at the scale 
of $B$ meson $\mu={\cal O}(m_b)$.

In order to calculate the branching ratios and CP asymmetries in this work, we use the next 
to leading order Wilson coefficients for $\Delta B=1$ transitions obtained in the naive 
dimensional regularization scheme (NDR) at the energy scale $\mu=m_b(m_b)$, 
$\Lambda^{(5)}_{\overline{MS}}=225$ MeV and quark top mass $m_t=170$ GeV. These coefficients are 
taken from Ref. \cite{buras96}, see Table XXII. Those values are $c_1=1.082$, $c_2=-0.185$, 
$c_3=0.014$, $c_4=-0.035$, $c_5=0.009$, $c_6=-0.041$, $c_7/\alpha=-0.002$, 
$c_8/\alpha=0.054$, $c_9/\alpha=-1.292$ and $c_{10}/\alpha=0.263$, where 
$\alpha=1/137$ is the fine structure constant.

\subsection{Generalized naive factorization approach}

The decay amplitude of a nonleptonic two body $B$ decay can be calculated using the
effective weak Hamiltonian by

\ba {\cal M}(B\to M_1 M_2) = \langle M_1 M_2|H_{eff}|B\rangle = 
{G_F\over \sqrt{2}}\sum^{10}_{i=1} C_i(\mu) \langle O_i(\mu)\rangle\ ,\ea

\noindent where the hadronic matrix elements $\langle O_i(\mu)\rangle$ are 
defined by $\langle M_1 M_2|O_i(\mu)|B\rangle$ and $M_i$ are final state mesons. 
In the naive factorization hypothesis, hadronic matrix elements $\langle O_i(\mu)\rangle$ 
are evaluated by the product of decay constants and form factors.
These matrix elements are energy $\mu$ scale and renormalization scheme independent,
consequently there is no term to cancel the energy $\mu$ dependency in the Wilson 
coefficients, and the amplitudes for nonleptonic two body $B$ decays are scale and 
renormalization scheme dependent.

The improved naive factorization approach \cite{ali99, ali} is formulated to solve the problem 
of energy scale dependency by including some perturbative QCD contributions in Wilson 
coefficients. This is considered in order to isolate the energy $\mu$ dependency from the 
matrix element $\langle O_i(\mu)\rangle$ and join it with the Wilson coefficients to produce 
effective Wilson coefficients $c^{eff}_i$, which are scale $\mu$ independent. Schematically 

\ba \sum_i C_i(\mu)\langle O_i(\mu)\rangle =
\sum_i C_i(\mu)g_i(\mu)\langle O_i\rangle_{tree} =
\sum_i c^{eff}_i\langle O_i\rangle_{tree}\ ,\ea

\noindent where $g_i(\mu)$ are perturbative QCD corrections to Wilson coefficients
and $\langle O_i\rangle_{tree}$ are tree level hadronic matrix elements.
Explicit expressions for the effective Wilson coefficients $c^{eff}_i$ are given in 
Refs. \cite{ali}. These coefficients are recalculated with the current CKM parameters 
\cite{pdg2008}. Effective Wilson coefficients, for transitions $b\to d$ and $b\to s$,
are shown in Table I. They are evaluated at the factorisable scale $\mu=m_b$, with an 
averaged momentum transfer of $k^2=m^2_b/2$, and using the central values for CKM parameters 
from Ref. \cite{pdg2008}.

\begin{table}[ht]
{\small TABLE I.~Effective Wilson coefficients $c^{eff}_i$ for $b\to d$ and $b\to s$
transitions. Evaluated at $\mu_f=m_b$ and $k^2=m^2_b/2$, where the central values for 
Wolfenstein parameters $\lambda=0.2257$, $A=0.814$, $\rho=0.135$ and $\eta=0.349$ are used, 
see Ref. \cite{pdg2008}}
\begin{center}
\begin{tabular}{l c c}
\hline \hline
$c^{eff}_i$           & $b\to d$           & $b\to s$ \\
\hline
$c^{eff}_1$           &  1.1680            &  1.1680           \\
$c^{eff}_2$           & -0.3652            & -0.3652            \\
$c^{eff}_3$           &  0.0231 + i 0.0038 &  0.0233 + i 0.0043 \\
$c^{eff}_4$           & -0.0477 - i 0.0113 & -0.0482 - i 0.0129 \\
$c^{eff}_5$           &  0.0139 + i 0.0038 &  0.0140 + i 0.0043 \\
$c^{eff}_6$           & -0.0499 - i 0.0113 & -0.0503 - i 0.0129 \\
$c^{eff}_7/\alpha$    & -0.0303 - i 0.0326 & -0.0311 - i 0.0356 \\
$c^{eff}_8/\alpha$    &  0.0551            &  0.0551            \\
$c^{eff}_9/\alpha$    & -1.4268 - i 0.0326 & -1.4276 - i 0.0356 \\
$c^{eff}_{10}/\alpha$ &  0.4804            &  0.4804            \\
\hline \hline
\end{tabular}
\end{center}
\end{table}

In the factorisable decay amplitude, the effective Wilson coefficients appear 
as linear combinations. Thus, to simplify decay amplitudes, $a_i$ coefficients are 
introduced

\ba a_i &\equiv& c^{eff}_i + \frac{1}{N_c} c^{eff}_{i+1}\ (i=odd)\ ,\nn\\
a_i &\equiv& c^{eff}_i + \frac{1}{N_c} c^{eff}_{i-1}\ (i=even)\ ,\ea

\noindent where index $i$ runs over $1,...,10$ and $N_c=3$ is the color number of QCD.
Effective coefficients $a_i$ for $b\to d$ and $b\to s$ transitions are shown in Table II.

\begin{table}[ht]
{\small TABLE II.~Effective coefficients $a_i$ for $b\to d$ and $b\to s$ transitions
(in units of $10^{-4}$ for $a_3$, ..., $a_{10}$).}
\begin{center}
\begin{tabular}{l c c}
\hline \hline
$a_i$    & $b\to d$        &  $b\to s$     \\
\hline
$a_1$    & 1.046           &  1.046         \\
$a_2$    & 0.024           &  0.024         \\
$a_3$    & 72              &  72            \\
$a_4$    & -400   - i 101  & -404   - i 114 \\
$a_5$    & -28             & -28            \\
$a_6$    & -453   - i 101  & -457   - i 114 \\
$a_7$    & -0.87  - i 2.38 & -0.93  - i 2.60 \\
$a_8$    &  3.28  - i 0.79 &  3.26  - i 0.87 \\
$a_9$    & -92.5  - i 2.38 & -92.5  - i 2.60 \\
$a_{10}$ &  0.35  - i 0.79 &  0.33  - i 0.87 \\
\hline \hline
\end{tabular}
\end{center}
\end{table}

\section{Input parameters and form factors}

\subsection{Input parameters}

The CKM matrix is parametrized in terms of Wolfenstein parameters $\lambda$, $A$, 
$\bar \rho$ and $\bar \eta$ \cite{wolfenstein}, 

\ba \left(\begin{array}{ccc}
1-\frac{1}{2}\lambda^2 & \lambda & A\lambda^3(\rho-i\eta) \\
-\lambda & 1-\frac{1}{2}\lambda^2 & A\lambda^2 \\
A\lambda^3(1-\bar\rho-i\bar\eta) & -A\lambda^2 & 1
\end{array}\right) ,\ea

\noindent with $\bar\rho=\rho(1-\lambda^2/2)$ and $\bar\eta=\eta(1-\lambda^2/2)$,
including ${\cal O}(\lambda^5)$ corrections \cite{buras94}. The Wolfenstein parameters 
are determined by unitarity constrain of three family of quarks and a global fit to 
experimental data. The central values $\lambda=0.2257$, $A=0.814$, $\bar \rho=0.135$, 
and $\bar \eta=0.349$ are used in calculations, see Ref. \cite{pdg2008}.

Running quark masses enter in loop calculation of effective Wilson coefficients. 
Furthermore, they are present in the equation of motion necessary to calculate the chiral factor,
which multiply the matrix elements of penguin terms $a_6$ and $a_8$ in the effective weak 
Hamiltonian. These contributions are only present in processes involving pseudoscalar 
mesons $\pi$, $\eta$, $\eta'$, $K$ and $\pi'$ in final states.

Since the energy release in $B$ decay is of order $m_b$, the scale energy for evaluation of
running quark masses should be $\mu\approx m_b$. The values $m_u(m_b)=3.2$ MeV, 
$m_d(m_b)=6.4$ MeV, $m_s(m_b)=127$ MeV, $m_c(m_b)=0.95$ GeV and $m_b(m_b)=4.34$ GeV are used 
in calculations, see Ref. \cite{fusaoku}.

The decay constants of pseudoscalar and vector mesons are determined using branching ratio of 
mesons and $\tau$ semileptonic decays, respectively. The central values, 
$f_\pi=130$ MeV, $f_K=160$ MeV, $f_\rho=212$ MeV, $f_\omega=195$ MeV, $f_{K^*}=221$ MeV and 
$f_{\phi}=237$ MeV are extracted using experimental data, see Ref. \cite{pdg2008}. 

In Ref. \cite{holl} is presented the argument that decay constants of 
radial excited pseudoscalar mesons are suppressed relative to the pion decay constant.
Decay constant $f_{\pi'}$ computed in models has also been found to be small, see Ref. \cite{michael}.
In unquenched lattice QCD \cite{michael}, the ratio $f_{\pi'}/f_{\pi}=0.078(93)$ is calculated.
Using $f_\pi=131$ MeV, the value of decay constant $f_{\pi'}=16.3\pm 6.1$ MeV is obtained. 
From the experimental bound on branching ratio $Br(\tau\to \pi'\nu_\tau)$ \cite{asner} 
is established a bound in the decay constant, $f_{\pi'}< 8.4$ MeV \cite{diehl}. 
In Ref. \cite{arndt99}, the authors obtain $f_{\pi'} = 26$ MeV and $f_{\rho'}=128$ MeV, using a light 
cone quark model. Recently, in the large-$N_c$ limit and using QCD spectral sum rules in Ref. \cite{cata}, 
the authors estimate the decay constant $f_{\rho'}=182\pm5$ MeV.

In order to calculate branching ratios and CP asymmetries for channels $B\to P\pi'$ and 
$B\to V\pi'$ two values of decay constant are used $f_{\pi'}=0$ MeV and $26$ MeV, which are 
the minimum and maximum in the range of values obtained by different methods. For channels 
$B\to P\rho'$ and $B\to V\rho'$ the decay constant $f_{\rho'}=0.128$ MeV is used to calculate 
branching ratios and CP asymmetries.

The mixing of the $\eta-\eta'$, $\rho^0-\omega$ and $\omega-\phi$ systems are considered
by mixing in decay constants and form factors. Ideal mixing is considered for $\omega-\phi$, 
i.e. the mesons have quark content $\omega=1/\sqrt{2}(u\bar u+d\bar d)$ and $\phi=s\bar s$.

Two mixing angle formalism is used to describe mixing in $\eta-\eta'$ system, see 
Refs. \cite{leutwyler98,feldmann98}. The physical states $\eta$ and $\eta'$ are defined
in terms of flavor octet $\eta_8$ and singlet $\eta_0$. 
In Ref. \cite{feldmann98}, we find a complete fit of mixing parameters to 
experimental data, resulting for decay constants the following central values 
$f^u_\eta=76.2$ MeV, $f^u_{\eta'}=61.8$ MeV, $f^s_\eta=-110.5$ MeV and 
$f^s_{\eta^\prime}=138$ MeV. Decay constants $f^c_\eta = -(2.4\pm 0.2)$ MeV and 
$f^c_{\eta'}=-(6.3\pm 0.6)$ MeV are used to include the $\eta_c$ meson in the mixing scheme.
When scalar and pseudoscalar densities in penguin terms $a_6$ and $a_8$ are evaluated, the
correct chiral behavior must be ensured. Thus, these matrix elements are multiplied by the 
factor $r_{\etap}$. The numerical values $r_\eta=-0.689$ and $r_{\eta^\prime}=0.462$ are used,
see Ref. \cite{cheng98}.

For the meson $B$ lifetime, the values $\tau_{B^-}=(1.638\pm 0.011)\times 10^{-6}$ s and
$\tau_{\bar B^0}=(1.530\pm 0.009)\times 10^{-6}$ s are used, and for $B$ mass 
$m_{B^-}=5279.15\pm 0.31$ MeV and $m_{\bar B^0}=5279.53\pm 0.31$ MeV \cite{pdg2008}, 
which are required to calculate branching ratios.

\subsection{Form factors}

The WSB model \cite{wsb} and LCSR approach \cite{ball} are used to calculate form factors for 
$B\to P$ and $B\to V$ transitions. Since the WSB model and LCSR approach provide form factors
only for the above transitions, form factors for $B\to \pi'$ and $B\to \rho'$ transitions are 
calculated using the ISGW2 quark model \cite{isgw95}.

The transitions $B\to P$ and $B\to V$ can be written in terms of form factors by the 
following expressions

\ba \langle P(p_P)|V_\mu| B(p_B)\rangle &\equiv& \Bigg[(p_B+p_P)_\mu
-\frac{m^2_B-m^2_P}{q^2}\ q_\mu\Bigg]F_1(q^2)
+\Bigg[\frac{m^2_B-m^2_P}{q^2}\Bigg]q_\mu F_0(q^2) \ea

\noindent and

\ba \langle V(p_V,\epsilon)|(V_\mu-A_\mu)| B(p_B)\rangle &\equiv&
-\epsilon_{\mu \nu \alpha \beta}\epsilon^{\nu *}p^\alpha_B p^\beta_V
\frac{2V(q^2)}{(m_B+m_V)}-i\Bigg[\left(\epsilon^*_\mu-
\frac{\epsilon^*\cdot q}{q^2}q_\mu\right)(m_B+m_V)A_1(q^2)\\
&&-\left((p_B+p_V)_\mu-\frac{(m^2_B-m^2_V)}{q^2}q_\mu\right)(\epsilon^*\cdot q) 
\frac{A_2(q^2)}{(m_B+m_V)} + \frac{2m_V(\epsilon^*\cdot q)}
{q^2}q_\mu A_0(q^2) \Bigg] \label{ffvector}\nn\ea

\noindent where $q=(p_B-p_P)$ or $q=(p_B-p_V)$ and $\epsilon$ is the polarization of the vector 
meson $V$. The following restrictions are imposed over form factors in order to cancel poles at
$q^2=0$

\ba F_1(0) &=& F_0(0) ,\nn\\
2m_V A_0(0) &=& (m_B+m_V)A_1(0)-(m_B-m_V)A_2(0) .\ea

In the case of the WSB model a single pole dominance model is used for the $q^2$ 
momentum squared dependency

\be f(q^2)=\,\frac{f(0)}{\left(1-{q^2/m^2_*}\right)} ,\ee

\noindent where $m^2_*$ is the pole mass given by the vector meson and $f(0)$ is 
the form factor at zero momentum transfer.

The WSB model is a relativistic constituent quark model where the meson-meson matrix
elements are evaluated from the average integral corresponding to meson functions, 
which are solutions of a relativistic harmonic oscillator potential \cite{wsb}.

The LCSR approach uses the method of QCD sum rules on the light cone \cite{ball}.
The second set of parameters are used in the calculations, see \cite{ball}. A fit 
parametrization is utilized for the $q^2$ dependency of the form factors.

In Table III, values for form factors involved in transitions $B\to P$ and
$B\to V$ are shown, evaluated at zero momentum transfer, in the WSB quark model \cite{wsb}
and LCSR approach \cite{ball}. Form factor in $B\to \pi$ transition, evaluated 
in the LCSR approach, is small compared to the WSB model. This result is in accordance with 
current experimental data \cite{pdg2008}. The pseudoscalar meson $\eta^\prime$ is too heavy 
to be treated in the LCSR approach, its value is not reported. In general, the form factors 
calculated in LCSR approach are smaller than those calculated in the WSB model, for this reason 
the branching ratios are also smaller.

\begin{table}[ht]
{\small TABLE III. ~Form factors at zero momentum transfer for $B\to P$ and $B\to V$
transitions, evaluated in the WSB quark model \cite{wsb} and LCSR approach \cite{ball}.}
\begin{center}
\begin{tabular}{l c c c c c} \hline \hline
Transition  & $F_1=F_0$ & $V$ & $A_1$ & $A_2$ & $A_0$ \\
\hline
$B\to\pi$    & 0.333 [0.258]& & & &  \\
$B\to K$     & 0.379 [0.331]& & & & \\
$B\to\eta$   & 0.168 [0.275]& & & & \\
$B\to\eta'$  & 0.114 [-]& & & & \\
$B\to\rho$   & & 0.329 [0.323]& 0.283 [0.242]& 0.283 [0.221]& 0.281 [0.303]\\
$B\to\omega$ & & 0.232 [0.311]& 0.199 [0.233]& 0.199 [0.181]& 0.198 [0.363]\\
$B\to K^*$   & & 0.369 [0.293]& 0.328 [0.219]& 0.331 [0.198]& 0.321 [0.281]\\
\hline \hline
\end{tabular}
\end{center}
\end{table}

The improved version of the ISGW model \cite{isgw89}, the so called ISGW2 model \cite{isgw95}, 
a non relativistic quark model is used in this work. Even tough, in the ISGW model is possible 
to calculate transitions to a radially excited pseudoscalar and vector mesons, the ISGW2 model
is better because the constrains imposed by heavy quark symmetry, hyperfine distortions of wave
functions, and form factor more realistic at high recoil momentum transfer. These additional 
features incorporated in the ISGW2 model allow us to make more reliable estimations.

In the ISGW2 model, $B\to P'$ transition is written as

\ba \langle P(p_P)|V_\mu| B(p_B)\rangle &\equiv& 
f^\prime_+(q^2) (p_B+p_P)_\mu + f^\prime_-(q^2)(p_B-p_P)_\mu\ea

\noindent and matrix elements of vector and axial vector currents for $B\to V^\prime$ 
transition are written as

\ba \langle V(p_V,\epsilon)|V_\mu| B(p_B)\rangle &\equiv&
i g^\prime(q^2) \epsilon_{\mu\nu\alpha\beta}\epsilon^{*\nu}
(p_B+p_V)^\alpha(p_B-p_V)^\beta \ea

\noindent and

\ba \langle V(p_V,\epsilon)|A_\mu| B(p_B)\rangle &\equiv&
f^\prime(q^2) \epsilon^*_\mu + a^\prime_+(q^2)(\epsilon^* \cdot p_B)(p_B+p_V)_\mu
+ a^\prime_-(q^2)(\epsilon^* \cdot p_B)(p_B-p_V)_\mu ,\ea

\noindent where $q=(p_B-p_P)$ or $(p_B-p_V)$ in the respective case.

Form factors in ISGW2 model are related to form factors in WSB model
by the following relations 

\ba F_1(q^2) &=& f^\prime_+(q^2), \nn\\
V(q^2) &=& (m_B+m_V)\, g^\prime(q^2), \nn \\
A_1(q^2) &=& (m_B+m_V)^{-1} f^\prime(q^2), \nn\\
A_2(q^2) &=& -(m_B+m_V)\ a^\prime_+(q^2), \nn\\
A_0(q^2) &=& {1\over 2 m_V}\left[f^\prime(q^2)+(m^2_B-m^2_V)\, a^\prime_+(q^2),
+ q^2\, a^\prime_-(q^2)\right] . \ea

The form factors for $B\to \pi'$ and $B\to \rho'$ transitions, calculated at momentum 
transfer $q^2=m^2_\pi$, are presented in Table IV. To estimate branching ratios,
it is necessary to calculate form factors at different momentum transfers, namely 
at $q^2=m^2_K$, $m^2_\eta$, $m^2_{\eta'}$, $m^2_{\omega}$, $m^2_{K^*}$ and $m^2_{\phi}$.

\begin{table}[ht]
{\small TABLE IV.~Form factors at momentum transfer $q^2=m^2_\pi$ for $B\to \pi'$ and 
$B\to \rho'$ transitions, evaluated in the ISGW2 quark model \cite{isgw95}.}
\begin{center}
\begin{tabular}{l c c c c c} 
\hline \hline
Transition  & $F_1=F_0$ & $V$ & $A_1$ & $A_2$ & $A_0$ \\
\hline
$B\to\pi'$  & 0.25 & & & &  \\
$B\to\rho'$ & & 0.456 & 0.118 & -0.118 & 0.397 \\
\hline \hline
\end{tabular}
\end{center}
\end{table}

Mixing $\eta-\eta'$ effect is not included in the WSB model prediction. $SU(3)$ symmetry is used 
to consider mixing in $B\to \eta$ and $B\to \eta'$ transitions, which imply the relations 
$F^{B\pi}(0)=\sqrt{3}F^{B\eta_0}(0)=\sqrt{6}F^{B\eta_8}(0)$ and

\ba F^{B\eta} &=& F^{B\eta_8}\cos\theta - F^{B\eta_0}\sin\theta\ ,\nn\\
F^{B\eta'} &=& F^{B\eta_8}\sin\theta + F^{B\eta_0}\cos\theta\ ,\ea

\noindent where $\theta=-15.4^\circ$ is the mixing angle \cite{feldmann98}. Using
$F^{B\pi}(0)=0.333$ from the WSB model, form factors $F^{B\eta}(0)=0.181$ and $F^{B\eta'}(0)=0.148$ 
are obtained. In the LCSR approach, using the form factor $F^{B\pi}(t)$ and $SU(3)$ symmetry,
the form factors $F^{B\eta}(t)$ and $F^{B\eta'}(t)$ are estimated.

The $\rho^0-\omega$ mixing is introduced in hadronic matrix element $B\to \rho^0$. 
Nevertheless, the effect in $B\to \omega$ transitions is negligible and it is not included 
in branching ratios calculations. In the limit of isospin symmetry, physical states $\rho^0$ 
and $\omega$ are expressed in terms of isospin eigenstates $\rho^I$ and $\omega^I$ by a rotation
matrix 

\ba |\rho^0\rangle &=& |\rho^I\rangle + \epsilon |\omega^I\rangle \nn\\
|\omega \rangle &=& |\omega^I\rangle - \epsilon'|\rho^I\rangle ,\ea

\noindent where numerical values for mixing parameters are $(1+\epsilon)=(0.092+0.016i)$ 
and $(1-\epsilon')=(1.011+0.030i)$. Including isospin effects, hadronic matrix element for 
$B\to \rho^0$ transition is modified by the factor $(1+\epsilon)$, see Ref. \cite{diaz}.

\section{Amplitudes and branching ratios}

The amplitudes for processes studied in this work are explicitly written in the
Appendices. These amplitudes are given in terms of decay constants and form factors, and 
contain all the contributions of the effective weak Hamiltonian.  

Appendix A contains the amplitudes for $B\to P\pi'$ decays, where $P$ is the pseudoscalar
meson $\pi$, $\eta$, $\eta'$ or $K$. The amplitude for the process $\bar B^0 \to \pi^- \pi'^-$ is 
not written, but it can be obtained directly from $\bar B^0 \to \pi^+ \pi'^+$, interchanging $\pi$ 
by $\pi'$. Similarly, the amplitude for $B^-\to \pi^-\pi'^0$ can be obtained 
from $B^-\to \pi^0 \pi'^-$.

To compare $B\to P\pi'$ with $B\to P\pi$ modes, besides obvious differences in decay constants and 
form factors, a point to remark is the following one. In the penguin sector the more 
important contributions come from terms $a_4$ and $a_6$. Particularly, the $a_6$ and $a_8$ coefficients 
are enhanced by a chiral factor, which is proportional to the squared mass of pseudoscalar $P$ or 
pseudoscalar radial excitation $\pi'$. In the case of radial excitation $\pi'$, this contribution
can be two orders of magnitude bigger than contribution of the pseudoscalar meson $\pi$. 
This enhancement effect is shown in the branching ratios of channels $B\to \pi\pi'$, 
$B\to \eta\pi'$ and $B\to \eta'\pi'$. 

In Appendix B, the amplitudes for $B\to V\pi'$ processes are shown, where $V$ is the vector
meson $\rho$, $\omega$, $K^*$ or $\phi$. In these modes, the increased factor in the $a_6$ 
and $a_8$ penguin contributions occur like in $B\to P\pi'$ modes. This effect appears in the 
branching ratios of modes $B\to \rho\pi'$ and $B\to \omega\pi'$, with the exception 
of channel $\bar B^0\to \rho^-\pi'^+$.

The amplitudes for $B\to P\rho'$ processes are given in Appendix C. The processes 
$B\to \pi\rho'$ are not written, but they can be obtained from amplitudes $B\to \rho\pi'$ 
in Appendix B, interchanging $\pi'$ by $\pi$ and $\rho$ by $\rho'$.

In Appendix D, the decay amplitudes for $B\to V\rho'$ processes are given by the factorized term 

\ba X^{B\rho^\prime, V} &=& \langle V|(\bar q_3 q_2)_{V-A}|0\rangle 
\langle\rho^\prime|(\bar q_1 b)_{V-A}|B \rangle \\
&=& -if_V m_V \left[(\epsilon^*_V\cdot \epsilon^*_{\rho^\prime})(m_B+m_{\rho^\prime})A^{B\rho^\prime}_1(m^2_V)
-(\epsilon^*_V\cdot p_B)(\epsilon^*_{\rho^\prime}\cdot p_B){2A^{B\rho^\prime}_2(m^2_V)
\over (m_B+m_{\rho^\prime})} +i\epsilon_{\mu\nu\alpha\beta}\epsilon^\mu_V \epsilon^\nu_{\rho^\prime}
p^\alpha_B p^\beta_{\rho^\prime}{2V^{B\rho^\prime}(m^2_V)\over (m_B+m_{\rho^\prime})} \right],\nn
\ea

\noindent where $\epsilon_V$ and $\epsilon_{\rho'}$ are the polarization vectors of the vectors mesons 
$V$ and $\rho'$, respectively. This notation is introduced to simplify expressions in Appendix D.

The amplitudes for processes which contain the meson $\pi^0$, $\pi'^0$, $\rho^0$ or $\rho'^0$ 
in the final state, are multiplied by the factor $1/ \sqrt{2}$ due to the wave function of these
neutral mesons, i.e. $1/ \sqrt{2}(\bar u u - \bar d d)$.

From decay amplitude and input parameters, branching ratios are straightforward calculated.  
The decay rate for processes $B\to P\pi'$ is given by

\ba \Gamma(B\to P\pi^\prime) &=& {\lambda^{1/2}(m^2_B,m^2_P,m^2_{\pi^\prime})\over 
16\pi\,  m^3_B} |{\cal M}(B\to P\pi^\prime)|^2 ,\label{br}\ea

\noindent where $\lambda(x,y,z)=x^2+y^2+z^2-2(xy+xz+yz)$. The decay rate for processes 
$B\to V\pi'$ and  $B\to P\rho'$ are calculated using Eq. \eqref{br}. In this case, the squared 
amplitude is proportional to $|\epsilon_V\cdot p_{\pi'}|^2$ and $|\epsilon_{\rho'}\cdot p_P|^2$, 
respectively. In processes $B\to V\rho'$, the squared amplitude is involved due to 
interfering terms proportional to $X^{B\rho', V}$ and $X^{BV, \rho'}$ contributions.

The branching ratios listed in Tables V and VI, are CP averaged conjugate modes. 
Charged and neutral channels are calculated by

\ba &&{\Gamma(B^+\to f^+) + \Gamma(B^-\to f^-)\over 2},\, \,
{\Gamma(B^0\to f) + \Gamma(\bar B^0\to \bar f)\over 2} ,\ea

\noindent where $f$ is the two body meson final state, i.e, $f=P\pi'$, $V\pi'$, $P\rho'$ 
or $V\rho'$.

\begin{table}[ht]
{\small Table V. Branching ratios (in units of $10^{-6}$) averaged over CP conjugate modes
for $B\to P\pi'$ and $B\to V\pi'$ decays, using the WSB \cite{wsb} model and 
LCSR approach \cite{ball}, decay constants $f_{\pi'}=26$ MeV and $f_{\pi'}=0.0$ MeV.}
\begin{center}
\renewcommand{\arraystretch}{1.2}
\renewcommand{\arrayrulewidth}{0.8pt}
\begin{tabular}{l c l c}
\hline \hline
Mode                       & ${\cal B}$  & Mode     &  ${\cal B}$  \\
\hline
$\bar B^0 \to \pi^+ \pi'^-$ & 5.8 [5.8] 5.8 & $\bar B^0 \to \rho^- \pi'^+$   & 14.6 [14.6] 14.6\\
$\bar B^0 \to \pi^- \pi'^+$ & 41.4 [26.7] 0.0 & $\bar B^0 \to \rho^+ \pi'^-$ & 23.2 [30.2] 0.0\\
$\bar B^0 \to \pi^0 \pi'^0$ & 6.7 [4.5] 0.03 & $\bar B^0 \to \rho^0 \pi'^0$  & 2.8  [3.8] 0.02\\
$B^- \to \pi^- \pi'^0$      & 6.0 [6.0] 3.0 & $B^- \to \rho^0 \pi'^-$        & 13.8 [17.7] 0.04\\
$B^- \to \pi^0 \pi'^-$      & 0.47 [0.29] 0.005 & $B^- \to \rho^- \pi'^0$    & 14.1 [16.1] 7.8\\
$\bar B^0 \to \eta \pi'^0$ & 4.2 [2.9] 0.05 & $\bar B^0 \to \omega \pi'^0$  & 7.1 [14.8] 0.05\\
$B^- \to \eta \pi'^-$      & 15.0 [10.1] 0.1 & $B^- \to \omega \pi'^-$      & 7.6 [24.1] 0.1\\
$\bar B^0 \to \eta' \pi'^0$ & 2.3 [1.6] 0.007 & $\bar B^0 \to K^{*-} \pi'^+$ & 3.0 [3.0]  3.0\\
$B^- \to \eta' \pi'^-$      & 8.6 [5.7] 0.02 & $\bar B^0 \to K^{*0} \pi'^0$  & 0.75 [0.75] 0.9\\
$\bar B^0 \to K^- \pi'^+$ & 6.5 [6.5] 6.5 & $B^- \to K^{*-} \pi'^0$     & 1.8 [1.8] 1.6\\
$\bar B^0 \to K^0 \pi'^0$ & 4.0 [4.0] 3.7 & $B^- \to \bar K^{*0} \pi'^-$ & 3.6 [3.6] 3.6\\
$B^- \to K^- \pi'^0$      & 3.8 [3.8] 3.5 & $\bar B^0 \to \phi \pi'^0$ & 0.004 [0.004] 0.004\\
$B^- \to \bar K^0 \pi'^-$ & 7.9 [7.9] 7.9 & $B^- \to \phi \pi'^-$      & 0.008 [0.008] 0.008\\
\hline \hline
\end{tabular}
\end{center}
\end{table}

\begin{table}[ht]
{\small Table VI. Branching ratios (in units of $10^{-6}$) averaged over CP conjugate modes
for $B\to P\rho'$ and $B\to V\rho'$ decays, using the WSB \cite{wsb} model and 
LCSR approach \cite{ball}, and decay constant $f_{\rho'}=128$ MeV.}
\begin{center}
\renewcommand{\arraystretch}{1.2}
\renewcommand{\arrayrulewidth}{0.8pt}
\begin{tabular}{lclc}
\hline \hline
Mode                       & ${\cal B}$    & Mode  &  ${\cal B}$ \\
\hline
$\bar B^0\to \pi^+ \rho'^-$ & 10.4 [6.8]  & $\bar B^0\to \rho^- \rho'^+$  & 38.7 [38.7] \\
$\bar B^0\to \pi^- \rho'^+$ & 13.2 [13.2] & $\bar B^0\to \rho^+ \rho'^-$  & 29.8 [22.5] \\
$\bar B^0\to \pi^0 \rho'^0$ & 0.02 [0.01] & $\bar B^0\to \rho^0 \rho'^0$  & 0.15 [0.13] \\
$B^-\to \pi^0 \rho'^-$      & 6.0 [4.0]   & $B^-\to \rho^0 \rho'^-$       & 7.9 [6.4] \\
$B^-\to \pi^- \rho'^0$      & 7.4 [7.4]   & $B^-\to \rho^- \rho'^0$       & 21.1 [21.1] \\
$\bar B^0\to \eta \rho'^0$  & 0.008 [0.006] & $\bar B^0\to \omega \rho'^0$ & 0.04 [0.03] \\
$B^-\to \eta \rho'^-$       & 3.4 [2.3]  & $B^-\to \omega \rho'^-$         & 4.1 [6.0] \\
$\bar B^0\to \eta' \rho'^0$ & 0.044 [0.039] & $\bar B^0\to K^{*-} \rho'^+$ & 8.0 [8.0] \\
$B^-\to \eta' \rho'^-$      & 2.1 [1.4]     & $\bar B^0\to K^{*0} \rho'^0$ & 6.6 [7.0] \\
$\bar B^0\to K^- \rho'^+$   & 1.1 [1.1]     & $B^-\to K^{*-} \rho'^0$      & 6.2 [6.6]\\
$\bar B^0\to K^0 \rho'^0$   & 0.6 [0.1]     & $B^-\to \bar K^{*0} \rho'^-$ & 9.5 [9.5] \\
$B^-\to K^- \rho'^0$        & 0.7 [0.6]     & $\bar B^0\to \phi \rho'^0$  & 0.01 [0.01] \\
$B^-\to \bar K^0 \rho'^-$   & 0.013 [0.013] & $B^-\to \phi \rho'^-$       & 0.02 [0.02] \\
\hline \hline
\end{tabular}
\end{center}
\end{table}

Form factors in transitions $B\to P$ and $B\to V$ are calculated using two representative methods, 
the WSB quark model \cite{wsb} and LCSR approach \cite{ball}. The branching ratios calculated in LCSR 
approach are listed between squared brackets in Tables V and VI. In transitions $B\to \pi'$ and $B\to \rho'$ 
the improved version of the ISGW non relativistic quark model \cite{isgw95} is used.

In Table V, branching ratios for processes $B\to P\pi'$ and $B\to V\pi'$ are shown, where $P$ is the 
pseudoscalar meson $\pi$, $\eta$, $\eta'$ or $K$ and $V$ is the vector meson $\rho$, $\omega$, 
$K^*$ or $\phi$. Branching ratios are calculated using two different values in decay constant 
$f_{\pi'}=26$ MeV (first and second column) and $f_{\pi'}=0$ MeV (third column).

In Table V, channels with $K$, $K^*$ or $\phi$ in the final state, have equal branching ratio calculated in 
the WSB model or in the LCSR approach. The branching ratio in processes $\bar B\to \pi^+\pi'^-$ and 
$\bar B\to \rho^-\pi'^+$ have independent value of the decay constant $f_{\pi'}$, since decay amplitude 
has only one contribution proportional to $f_{\pi}$ or $f_{\rho}$, respectively. 
Using the value $f_{\pi'}=26$ MeV in calculations,
channels $\bar B^0 \to \pi^- \pi'^+$, $B^- \to \eta \pi'^-$, $\bar B^0 \to \rho^- \rho'^+$, 
$\bar B^0 \to \rho^+ \rho'^-$ and $\bar B^0 \to K^{*-} \rho'^+$, have branching ratios of the order of 
$10^{-5}$. The channels with branching ratios of order below $10^{-6}$ are $B^-\to \pi^0\pi'^-$, 
$\bar B^0\to K^{*0}\pi'^0$, $\bar B^0\to \phi \pi'^0$ and $B^-\to \phi\pi'^-$.

Numerical values for branching ratios of processes $B\to P\rho'$ and $B\to V\rho'$ 
are listed in Table VI. The branching ratios are calculated using the decay constant $f_{\rho'}=128$ MeV. 
Branching ratio prediction for the decays $\bar B^0 \to \pi^- \rho'^+$, 
$\bar B^0 \to \rho^- \rho'^+$, $\bar B^0 \to \rho^+ \rho'^-$, $\bar B^0 \to K^{*-} \rho'^+$ are of 
order $10^{-5}$. The branching ratio of the channels $\bar B^0\to \pi^0\rho'^0$, 
$\bar B^0\to \etap \rho'^0$, $\bar B^0\to K^0\rho'^0$, $B^-\to K^-\rho'^0$ and 
$B^-\to \bar K^0\rho'^-$ are suppressed, of order below $10^{-6}$. The channels that include two 
vector mesons in final states $\bar B^0\to \rho^0\rho'^0$, $\bar B^0\to \omega\rho'^0$, 
$\bar B^0\to \phi\rho'^0$ and $B^-\to \phi\rho'^-$, have also branching ratios of order below $10^{-6}$. 
The modes $\bar B^0\to \pi^+\rho'^-$, $B^-\to \pi^0\rho'^-$, $B^-\to \eta \rho'^-$, 
$\bar B^0\to \rho^+\rho'^-$ have different branching ratios when form factors are calculated using the 
WSB model or the LCSR approach.

The authors in Ref. \cite{lipkin02} can calculate the ratios

\ba R_{\rho^+} = {Br(\bar B^0\to \pi^-\rho^{\prime +}) \over Br(\bar B^0\to \pi^-\rho^+)},\ \   
R_{\rho^0} = {Br(B^-\to \pi^-\rho^{\prime 0}) \over Br(B^-\to \pi^-\rho^0)} ,\ea

\noindent and obtain approximately $R_{\rho^+}\approx R_{\rho^0}\approx 2$. Using the world averaged 
experimental data from Ref. \cite{pdg2008},  $Br(\bar B^0\to \pi^-\rho^+)=10.9\pm 1.4\times 10^{-6}$ and 
$Br(B^-\to \pi^-\rho^0)=8.7\pm 1.1\times 10^{-6}$, it is possible to predict the branching ratios  
$Br(\bar B^0\to \pi^-\rho^{\prime +})=21.8\times 10^{-6}$ and 
$Br(B^-\to \pi^-\rho^{\prime 0})=17.4\times 10^{-6}$. These numbers are compared with our CP averaged 
branching ratios calculated in Table VI, $Br(\bar B^0\to \pi^-\rho^{\prime +})=13.2 \times 10^{-6}$ and 
$Br(B^-\to \pi^-\rho^{\prime 0})=7.4 \times 10^{-6}$.

Up to now, the only measured branching ratio reported is 
$Br(B^+\to \pi^+\rho(1450)^0)=1.4\pm0.4\pm0.4^{+0.3}_{-0.7} \times 10^{-6}$, see Refs. \cite{babar05, babar09}. 
Our estimate of this CP averaged branching ratio is $Br(B^-\to \pi^-\rho(1450)^0) = 7.4\times 10^{-6}$, 
see Table VI. This result is indicating that the decay constant $f_{\rho'}$ and the form factors in
the transitions $B\to \rho'$ can be overestimated.

\section{CP asymmetries}

Direct CP violation asymmetry in charged $B^\pm$ decays can be defined by  \cite{babarbook, bigibook}

\ba A_{CP} &=& {\Gamma(B^+\to f^+)-\Gamma(B^-\to f^-) \over 
\Gamma(B^+\to f^+)+\Gamma(B^-\to f^-)} . \ea

For $b\to q$ transitions (where $q=d,s$), the decay amplitudes can be written generically by

\ba {\cal M} &=& V_{ub}V^*_{uq}\, T - V_{tb}V^*_{tq}\, P\, ,\ea 

\noindent where $T$ is current-current contributions, $P$ is penguin QCD and electroweak
contributions, see the decay amplitudes in Appendices. 

In the standard model, CKM matrix elements contain weak phases due to the weak dynamics.
In the generalized naive factorization approach, the effective Wilson coefficients $c^{eff}_i(\mu)$ 
are complex numbers, which contain a strong phase due to QCD interactions \cite{aliCP}. Thus, phase 
contributions from the terms $T$ and $P$ in the decay amplitude can be factorized in terms of weak 
and strong phases. This result in the contributions $T'$ and $P'$. The decay amplitude ${\cal M}$ 
and its CP conjugate amplitude ${\cal \bar M}$ can be written as

\ba {\cal M} &=& e^{i\phi_1}e^{i\delta_1}\, T' 
+ e^{i\phi_2}e^{i\delta_2}\, P', \nn\\
{\cal \bar M} &=& e^{-i\phi_1}e^{i\delta_1}\, T' 
+ e^{-i\phi_2}e^{i\delta_2}\, P', \ea 

\noindent where $\phi_i$ weak decay phases change sign in a CP conjugate transformation and $\delta_i$ 
strong phases are conserved. Using this factorization of phases, direct CP violation asymmetry can be 
calculated by

\ba A_{CP} &=& {|{\cal M}|^2-|{\cal \bar M}|^2\over |{\cal M}|^2+|{\cal \bar M}|^2}
= {2\sin(\Delta \phi)\sin(\Delta \delta)\, r\over 1+r^2+2r\cos(\Delta \phi)\cos(\Delta \delta)} ,
\ea

\noindent where $r$ is the ratio $P'/T'$, $\Delta \phi=\phi_1-\phi_2$ and 
$\Delta \delta=\delta_1-\delta_2$ are the difference in the weak and the strong phase contributions 
to the terms $T'$ and $P'$. Three conditions must be fulfilled for the existence of direct CP 
violation. A CP violation weak phase $\Delta \phi$, final state interactions which induce a 
strong phase $\Delta \delta$, and two different contributions to the amplitude of comparable 
size $T' \approx P'$, see Refs. \cite{babarbook, bigibook}. 

In the framework of generalized naive factorization, the effective Wilson coefficients $c^{eff}_i$ 
are complex numbers. The imaginary part of these coefficients are due to calculable QCD perturbative 
contributions \cite{aliCP}. This effect induces a strong phase in the amplitudes, which is required 
to have direct CP violation. 

Direct CP violating asymmetries have been calculated using the form factors based on both the WSB \cite{wsb} 
model and LCSR approach \cite{ball}. The CP asymmetries depend weakly on the form factors. 
However, it is not the case for the weak decay constant $f_{\pi'}$. In the modes $B\to P\pi'$ and 
$B\to V\pi'$, dependency in the decay constant $f_{\pi'}$ will be discussed.

For the charged modes $B^\pm\to P\pi'$, $B^\pm\to V\pi'$, direct CP violating asymmetries are listed in 
Table VII. Table VII shows the dependency of direct CP asymmetries in the decay constant $f_{\pi'}$. 
Results using $f_{\pi'}=26$ MeV are shown in first and second columns, where the calculations
are done in the WSB model and LCSR approach, respectively. In third column is shown results using
$f_{\pi'}=0$ MeV. In this case, numerical results are independent of the model calculation of the form 
factors.

The processes $B^\pm \to K^\pm \pi'^0$, $B^\pm \to \rho^\pm \pi'^0$, $B^\pm \to K^{*\pm} \pi'^0$ have 
direct CP violating asymmetries of order $10\%$. The channels $B^\pm \to \bar K^0 \pi'^\pm$, 
$B^\pm \to \bar K^{*0} \pi'^\pm$ and $B^\pm \to \phi \pi'^\pm$ have only one contribution to its decay 
amplitude, in consequence direct CP asymmetry is equal to zero. Direct CP violating asymmetry 
in channels $B^\pm \to \pi^\pm \pi'^0$ and $B^\pm \to \omega \pi'^\pm$ depend on the use of WSB 
model or the LCSR approach.

In the channels $B^\pm \to \eta \pi'^\pm$, $B^\pm \to K^\pm \pi'^0$ and $B^\pm \to K^{*\pm} \pi'^0$,
direct CP violation asymmetries are not sensible to the value of decay constant $f_{\pi'}$.
When the value $f_{\pi'}=0$ MeV, direct CP violation asymmetry in channel $B^\pm \to \pi^\pm \pi'^0$ 
is equal to zero. In the same case, modes $B^\pm\to \pi^0\pi'^\pm$, $B^\pm\to \eta'\pi'^\pm$,  
$B^\pm\to \rho^0\pi'^\pm$ and $B^\pm\to \omega\pi'^\pm$ have an increase in the direct CP 
violation asymmetry. On the contrary, channel $B^\pm\to \rho^\pm\pi'^0$ has a decrease.

In Table VIII, direct CP violating asymmetries for the channels $B^\pm\to P\rho'$, $B^\pm\to V\rho'$ 
are shown, using the WSB model and LCSR approach, and the decay constant $f_{\rho'}=128$ MeV.

Direct CP violating asymmetry corresponding to the channels $B^\pm \to K^\pm \rho'^0$, 
$B^\pm \to \omega \rho'^\pm$ and $B^\pm \to K^{*\pm} \rho'^0$ are bigger than $10\%$, which make them good 
candidates to be observed experimentally. In channels $B^\pm\to \bar K^0\rho'^\pm$, 
$B^\pm\to \bar K^{*0}\rho'^\pm$ and $B^\pm \to \phi \rho'^\pm$, the decay amplitude has only one contribution 
in consequence the direct CP violating asymmetry is automatically equal to zero. The rest of channels have 
direct CP violating asymmetries of less than $10\%$ order. 

The channels $B^\pm \to \pi^0 \rho'^\pm$, $B^\pm \to \pi^\pm \rho'^0$ and $B^\pm \to \omega \rho'^\pm$ 
have direct CP violating asymmetries which depend on the use of the WSB model or the LCSR approach in 
evaluating the form factors in transitions $B\to \pi$ and $B\to \omega$.

\begin{table}[ht]
{\small Table VII. Direct CP violating asymmetries in percent for $B^\pm \to P\pi^\prime$ and
$B^\pm \to V\pi^\prime$ decays, using the WSB \cite{wsb} model and LCSR approach \cite{ball},
decay constants $f_{\pi'}=26$ MeV and $f_{\pi'}=0.0$ MeV.}
\begin{center}
\renewcommand{\arraystretch}{1.2}
\renewcommand{\arrayrulewidth}{0.8pt}
\begin{tabular}{l c l c}
\hline \hline
Final state                 & $A_{CP}$  & Final state              & $A_{CP}$ \\
\hline
$\pi^\pm \pi'^0$      & -1.4 [-1.1] 0.0& $\rho^0 \pi'^\pm$       & -5.6 [-5.5] -20.6\\
$\pi^0 \pi'^\pm$      & -0.4 [-0.5] 1.6& $\rho^\pm \pi'^0$       & -20.7 [-21.0] 4.8\\
$\eta \pi'^\pm$       & 4.8 [4.8] 4.6& $\omega \pi'^\pm$        & -6.3 [-5.8] 13.4\\
$\eta' \pi'^\pm$ & 5.3 [5.4] 20.3& $K^{*\pm} \pi'^0$        & -25.4 [-24.9] -27.7\\
$K^\pm \pi'^0$         & -12.5 [-12.6] -13.6& $\bar K^{*0} \pi'^\pm$ & 0.0 [0.0] 0.0\\
$\bar K^0 \pi'^\pm$    & 0.0 [0.0] 0.0& $\phi \pi'^\pm$          & 0.0 [0.0] 0.0\\
\hline \hline
\end{tabular}
\end{center}
\end{table}

\begin{table}[ht]
{\small Table VIII. Direct CP violating asymmetries in percent for $B^\pm \to P\rho'$ and 
$B^\pm \to V\rho'$ decays, using the WSB \cite{wsb} model and LCSR approach \cite{ball},
and decay constant $f_{\rho'}=128$ MeV.}
\begin{center}
\renewcommand{\arraystretch}{1.2}
\renewcommand{\arrayrulewidth}{0.8pt}
\begin{tabular}{lclc}
\hline \hline
Final state                 & $A_{CP}$ & Final state  & $A_{CP}$ \\
\hline
$\pi^0 \rho'^\pm$       & -3.9 [-5.8]&  $\rho^0 \rho'^\pm$   & 0.38 [0.38]\\
$\pi^\pm \rho'^0$       & 5.3  [6.0]&  $\rho^\pm \rho'^0$    & 0.39 [0.39]\\
$\eta \rho'^\pm$        & 5.2  [5.3]&  $\omega \rho'^\pm$    & 17.5 [15.2]\\
$\eta' \rho'^\pm$  & 5.0 [5.0]&  $K^{*\pm} \rho'^0$     & -20.2 [-19.4]\\
$K^\pm \rho'^0$         & -21.1 [-21.0]&  $\bar K^{*0} \rho'^\pm$ & 0.0 [0.0] \\
$\bar K^0 \rho'^\pm$    & 0.0 [0.0]&  $\phi \rho'^\pm$         & 0.0 [0.0]\\
\hline \hline
\end{tabular}
\end{center}
\end{table}

In neutral $B^0$ decays, because of the $B^0-\bar B^0$ mixing, it is required to include time dependent 
measurements in CP violation asymmetries. The CP violation time dependent asymmetry is defined as 

\ba A_f(t) &=& {\Gamma(B^0(t)\to f)-\Gamma(\bar B^0(t)\to \bar f) \over 
\Gamma(B^0(t)\to f)+\Gamma(\bar B^0(t)\to \bar f)}\nn\\
&=& C_f\cos(\Delta m \, t) + S_f \sin(\Delta m \, t) ,\ea

\noindent where $f$ is a two body final state. The coefficients $C_f$ and $S_f$ are defined by 

\ba C_f &=& {1-|\lambda_f|^2\over 1+|\lambda_f|^2} ,\,
S_f = {-2Im(\lambda_f)\over 1+|\lambda_f|^2} ,\ea

\noindent given in terms of the ratio $\lambda_f$, which is defined by

\ba \lambda_f &=& {V^*_{tb}V_{td}\langle f|H_{eff}|\bar B^0 \rangle \over
V_{tb}V^*_{td}\langle \bar f|H_{eff}|B^0 \rangle} .\ea

The coefficients $C_f$ and $S_f$ are functions of $\lambda_f$. The quantity $\lambda_f$ is 
independent of phase conventions and physically meaningful, in consequence the coefficients
$C_f$ and $S_f$ are observables. CP violation in the interference of decays with and without 
mixing is encoded in the coefficient $S_f\ne 0$. CP violation in decays means $C_f\ne 0$.

If the final state $f$ is a CP eigenstate, i.e. $CP |f\rangle= \pm |f\rangle$, and decay 
amplitudes are dominated by only one weak phase term contribution, then 
$\langle f|H_{eff}|\bar B^0 \rangle = \langle \bar f|H_{eff}|B^0 \rangle$,
 $C_f=0$ and $S_f=\eta_f \sin(2\phi)$, where $\eta_f$ is the CP eigenvalue of $f$ and
$2\phi$ is the difference in weak phase between the $B^0\to f$ and $B^0\to \bar B^0\to f$ 
decay path. A contribution of another term to the decay amplitude with a different weak phase make 
the value of $S_f$ depends on the strong phase. In this situation is also possible that $C_f\ne 0$. 

\begin{table}[ht]
{\small Table IX. CP violating asymmetry parameters $C_f$ and $S_f$ in percent for neutral 
$B^0(\bar B^0)\to P\pi'$ and $B^0(\bar B^0)\to V\pi'$ decays, using the WSB \cite{wsb} model 
and LCSR approach \cite{ball}, decay constants $f_{\pi'}=26$ MeV and $f_{\pi'}=0$ MeV.}
\begin{center}
\renewcommand{\arraystretch}{1.2}
\renewcommand{\arrayrulewidth}{0.8pt}
\begin{tabular}{l c c l c c}
\hline \hline
Final state              & $C_f$ & $S_f$ & Final state         & $C_f$ & $S_f$  \\
\hline
$\pi^0 \pi'^0$ & -0.7 [0.7] -9.0& -2.7 [-3.1] -30.5& $\rho^0 \pi'^0$& 1.2 [1.1] -20.6& 5.8 [4.7] -58.0\\
$\eta  \pi'^0$ & 0.4 [0.5] 4.6& 1.6 [2.1] 21.1& $\omega \pi'^0$ & 0.8 [0.5] 13.4 & 3.6 [2.4] 46.6\\
$\eta' \pi'^0$ & 0.5 [0.6] 20.3& 1.9 [2.5] 40.8& $K^{*\mp} \pi'^\pm$ & -27.7 [-27.7] -27.7& -17.7 [-17.7] -17.7\\
$K^\mp \pi'^\pm$ & -13.6 [13.6] 13.6& 30.1 [30.1] 30.1& $K^{*0} \pi'^0$ & 0.09 [0.12] 0.0 & 71.1 [71.1] 71.1\\
$K^0 \pi'^0$    & -0.1 [-0.1] -0.1& 70.7 [70.7] 70.7& $\phi \pi'^0$   & 0.0 [0] 0& 0.0 [0] 0\\
\hline \hline
\end{tabular}
\end{center}
\end{table}

\begin{table}[ht]
{\small Table X. CP violating asymmetries parameters $C_f$ and $S_f$ in percent for neutral 
$B^0(\bar B^0)\to P\rho'$ and $B^0(\bar B^0)\to V\rho'$ decays, using the WSB \cite{wsb} model 
and LCSR approach \cite{ball}, and decay constant $f_{\rho'}=128$ MeV.}
\begin{center}
\renewcommand{\arraystretch}{1.2}
\renewcommand{\arrayrulewidth}{0.8pt}
\begin{tabular}{l c c l c c}
\hline \hline
Final state                & $C_f$    & $S_f$ &Final state  & $C_f$ & $S_f$ \\
\hline
$\pi^0 \rho'^0$      & -48.8 [-53.9]& -83.7 [-83.3]& $\rho^0 \rho'^0$  & -20.6 [-20.6]& -58.0 [-58.0]\\
$\eta \rho'^0$       & -40.1 [-41.0]& -91.6 [-89.3]& $\omega \rho'^0$  & 20.0 [20.6]& 46.6 [46.6]\\
$\eta' \rho'^0$ & -5.1 [-4.3]& -48.4 [-47.6] & $K^{*\mp} \rho'^\pm$ &-27.7 [-27.7] &-17.8 [-17.8]\\
$K^\mp \rho'^\pm$    & -14.96 [-15.0]& -16.0 [-16.0]& $K^{*0} \rho'^0$ & -0.3 [-0.3]& 66.3 [66.3]\\
$K^0 \rho'^0$        & 0.8 [0.9]& 65.7 [65.7]& $\phi \rho'^0$ & 0.0 [0.0]& 0.0 [0.0]\\
\hline \hline
\end{tabular}
\end{center}
\end{table} 

CP violating asymmetry coefficients $C_f$ and $S_f$ for neutral $B^0(\bar B^0)$ decays with radial excited 
mesons $\pi'$ and $\rho'$ in final state are shown in Tables IX and X, respectively. For the processes 
$B^0(\bar B^0)\to \phi\pi'$ and $B^0(\bar B^0)\to \phi\rho'$ the coefficients $C_f$ and $S_f$ are equal to 
zero, since there is only one contribution to the decay amplitude in the respective channels.

The calculations results for the coefficients $C_f$ and $S_f$ are practically equal when the form factors
are estimated using the WSB model or the LCSR approach. Nevertheless, for the modes $B\to P\pi'$
and $B\to V\pi'$ there are a dependency with respect to use the values for the decay constant 
$f_{\pi'}=26$ MeV or $f_{\pi'}=0$ MeV. The channels with a strange meson in the final state have
the same value of the coefficients using the two different values in decay constant $f_{\pi'}$.

The channels with $C_f \approx 0$ and $S_f\ne 0$ are $B^0\to K^0\pi'^0$, $B^0\to K^{*0}\pi'^0$,
$B^0\to K^0\rho'^0$ and $B^0\to K^{*0}\rho'^0$. In these channels, where there is only present CP violation 
in the interference of the decay and in the mixing, it is possible to relate the coefficient $S_f$ to 
fundamental parameters in the standard model, i.e., interior angles of the unitary triangle.
 
\begin{table}[ht]
{\small Table XI. CP violating asymmetry parameters $\bar C_f$, $\bar S_f$, $\bar C_{\bar f}$ and 
$\bar S_{\bar f}$ in percent for $B^0(\bar B^0)\to \pi\pi'$, $B^0(\bar B^0)\to \rho\pi'$, 
$B^0(\bar B^0)\to \pi\rho'$ and $B^0(\bar B^0)\to \rho\rho'$ decays, using the WSB \cite{wsb} model and 
LCSR approach \cite{ball}.}
\begin{center}
\renewcommand{\arraystretch}{1.2}
\renewcommand{\arrayrulewidth}{0.8pt}
\begin{tabular}{l c c c c}
\hline \hline
Final states                  & $\bar C_f$ & $\bar S_f$ & $\bar C_{\bar f}$ & $\bar S_{\bar f}$ \\
\hline
$\pi^+ \pi'^-, \pi^- \pi'^+$    & 78.3 [68.2]& -72.3 [-60.0]& 61.5 [72.3]& 66.5 [77.0]\\
$\rho^+ \pi'^-,\rho^- \pi'^+$   & 22.7 [34.6]& -22.7 [-34.7]& -97.3[-93.8]& -89.8[-86.5]\\
$\pi^+ \rho'^-, \pi^- \rho'^+$  & 14.2 [34.0]& -9.1  [-29.3]& 4.2 [4.0]& 3.9 [3.8]\\
$\rho^+ \rho'^-,\rho^- \rho'^+$ & 15.4 [15.4]& -5.8  [-5.8]& 13.9 [13.9]& 14.0 [14.0]  \\
\hline \hline
\end{tabular}
\end{center}
\end{table}
 
CP violation in neutral $B^0$($\bar B^0$) mesons is involved in case that a final state $f$ and its CP
conjugate transformation state $\bar f$ are both common final states of $B^0$ and $\bar B^0$ mesons.
The final states $f$ and $\bar f$ are not CP eigenstates, i.e. $CP|f\rangle \ne |\bar f\rangle$.
For this case, time evolutions of the four decays $B^0(t)\to f$, $B^0(t)\to \bar f$, $\bar B^0(t)\to \bar f$,
$B^0(t)\to \bar f$, and $\bar B^0(t)\to f$ are studied in terms of four basic matrix elements

\ba g&=&\langle f|H_{eff}|B^0 \rangle,\, h=\langle f|H_{eff}|\bar B^0 \rangle, \nn\\ 
\bar g&=&\langle \bar f|H_{eff}|\bar B^0 \rangle,\, \bar h=\langle \bar f|H_{eff}|B^0 \rangle .\ea 

The following two CP violating asymmetries are introduced

\ba \bar A_f(t) &=& {\Gamma(B^0(t)\to f)-\Gamma(\bar B^0(t)\to f) \over 
\Gamma(B^0(t)\to f)+\Gamma(\bar B^0(t)\to f)}\nn\\
&=& \bar C_f\cos(\Delta m \, t) + \bar S_f \sin(\Delta m \, t)\ea

\noindent and

\ba \bar A_{\bar f}(t) &=& {\Gamma(B^0(t)\to \bar f)-\Gamma(\bar B^0(t)\to \bar f) \over 
\Gamma(B^0(t)\to \bar f)+\Gamma(\bar B^0(t)\to \bar f)}\nn\\
&=& \bar C_{\bar f}\cos(\Delta m \, t) + \bar S_{\bar f} \sin(\Delta m \, t) ,\ea

\noindent where the coefficients of $\cos(\Delta m\, t)$ and $\sin(\Delta m \, t)$ are defined by
 
\ba \bar C_f = {|g|^2-|h|^2 \over |g|^2+|h|^2}, \, 
\bar S_f= {-2Im({V^*_{tb}V_{td}\over V_{tb}V^*_{td}}{h\over g}) \over 1+|h/g|^2} ,\ea

\noindent and

\ba \bar C_{\bar f} = {|\bar h|^2-|\bar g|^2 \over |\bar h|^2+|\bar g|^2}, \,
\bar S_{\bar f} = {-2Im({V^*_{tb}V_{td}\over V_{tb}V^*_{td}}
{\bar g\over \bar h}) \over 1+|\bar g/\bar h|^2} .\ea

The condition for CP violation is that width decays $\Gamma(B^0(t)\to f)\ne \Gamma(\bar B^0\to f)$ 
and $\Gamma(B^0(t)\to \bar f)\ne \Gamma(\bar B^0\to \bar f)$, which means in terms of CP violating
asymmetry coefficients $\bar C_f\ne -\bar C_{\bar f}$ and (or) $\bar S_f\ne -\bar S_{\bar f}$.

Numerical values in percent for the CP violating asymmetry parameters $\bar C_f$, $\bar S_f$, 
$\bar C_{\bar f}$ and $\bar S_{\bar f}$  in the $B^0(\bar B^0)\to \pi\pi'$, $B^0(\bar B^0)\to \rho\pi'$, 
$B^0(\bar B^0)\to \pi\rho'$ and $B^0(\bar B^0)\to \rho\rho'$ decays, are listed in Table IX.
The form factors for the transitions $B\to \pi$ and $B\to \rho$ are calculated using the WSB model 
and the LCSR approach. 

The CP violating asymmetry parameters for the final states $\rho^+ \rho'^-,\rho^- \rho'^+$
have the same value if they are calculated using the WSB model or the LCSR approach. 
The parameters for the final states with a $\pi'$ mesons are only calculated using the
values in decay constant $f_{\pi}=26$ MeV. When the value $f_{\pi'}=0$ MeV is used, results are not
reported, since there are zero contributions, $g=0$ and $cg=0$, with the consequence that 
$\bar C_f=-\bar C_{\bar f}=100\%$, $\bar S_f= \infty$ and $\bar S_{\bar f}= 0$.

There is not yet measurements for CP violation asymmetries in the channels studied in this 
work \cite{pdg2008}. One of the reasons for this work is to estimate this asymmetries and
to motivate the experimental measurement of them.

\section{Conclusions}

In the framework of generalized naive factorization we calculate branching ratios and CP violating 
asymmetries of exclusive nonleptonic two body $B$ decays including the radial excited $\pi(1300)$ 
or $\rho(1450)$ meson in the final state. Branching ratios and CP violating asymmetries for the exclusive 
channels $B\to P\pi'$, $B\to V\pi'$, $B\to P\rho'$ and $B\to V\rho'$ (where, P and V denote a pseudoscalar 
and vector meson, respectively) have been estimated using all the contributions coming from the effective 
weak Hamiltonian $H_{eff}$. 

The form factors in $B\to P$ and $B\to V$ transitions are estimated using the WSB model \cite{wsb} and 
the LCSR approach \cite{ball}. In order to obtain form factors in $B\to \pi'$ and $B\to \rho'$ transitions, 
we use the improved version of the nonrelativistic ISGW quark model \cite{isgw89}, called ISGW2 model 
\cite{isgw95}. The factorized decay amplitudes for these decays are listed in Appendices.

We have obtained branching ratios for 52 exclusive channels. Some of these decays can be reached in 
experiments. In fact, decays $\bar B^0\to \pi^-\pi'^+$, $B^-\to \eta\pi'^-$, $\bar B^0\to \rho^-\pi'^+$,
$\bar B^0\to \rho^+\pi'^-$, $B^-\to \rho^0\pi'^-$, $\bar B^0\to \pi^-\rho'^+$, $\bar B^0\to \rho^-\rho'^+$,
$\bar B^0\to \rho^+\rho'^-$, and $B^-\to \rho^-\rho'^0$ have branching ratios of the order of $10^{-5}$.

We also studied the dependency of branching ratios in channels $B\to P\pi'$ and $B\to V\pi'$ with respect 
to the decay constant $f_{\pi'}$. The more sensible modes to the value in decay constant $f_{\pi'}$ are
$\bar B^0\to \pi^-\pi'^+$, $\bar B^0\to \pi^0\pi'^0$, $B^-\to \eta\pi'^-$, $B^-\to \eta'\pi'^-$, 
$\bar B^0\to \rho^+\pi'^-$, $B^-\to \rho^0\pi'^-$, $\bar B^0\to \omega\pi'^0$, and $B^-\to \omega\pi'^-$.
These channels could be the best scenario to determine the decay constant $f_{\pi'}$ in nonleptonic two 
body $B$ decays.

In general, we can explain the large branching ratios in decays $\bar B^0\to \pi^-\pi'^+$, 
$\bar B^0\to \rho^-\pi'^+$, $\bar B^0\to \rho^+\pi'^-$, and $B^-\to \rho^-\pi'^0$ by the effect of the 
enhancement of the chiral factor that multiply the penguin contributions $a_6$ and $a_8$
in the effective weak Hamiltonian $H_{eff}$.  

Direct CP violating asymmetry in channels $B^\pm\to K^\pm\pi'^0$, $B^\pm\to \rho^\pm\pi'^0$, 
$B^\pm\to K^{*\pm}\pi'^0$, $B^\pm\to \rho'^0$, $B^\pm\to \omega\rho'^\pm$, and $B^\pm\to K^{*\pm}\rho'^0$ 
are more than $10\%$ order. In the modes $B^\pm\to \eta'\pi^\pm$, $B^\pm\to \rho^0\pi'^\pm$, and 
$B^\pm\to \omega\pi'^\pm$, estimation of direct CP violating asymmetry using the value of the decay 
constant $f_{\pi'}=0$ MeV, give an increase with respect to the calculations using the value $f_{\pi'}=26$ MeV. 
On the contrary the channel $B^\pm\to \rho^\pm\pi'^0$ has a decrease in its estimation.
When the value in the decay constant $f_{\pi'}=0$ MeV is used, estimation of direct CP violating 
asymmetry in modes $B^\pm\to \eta'\pi'^\pm$, $B^\pm\to \rho^0\pi'^\pm$, and $B^\pm\to \omega\pi'^\pm$ 
are more than $10\%$ order.

In the neutral modes $B^0\to K^0\pi'^0$, $B^0\to K^{*0}\pi'^0$, $B^0\to K^0\rho'^0$ and 
$B^0\to K^{*0}\rho'^0$, we have estimated the CP violating asymmetry coefficientes $C_f\approx 0$ and 
$S_f$ more than $60\%$. 

Finally, we want to mention that our predictions for the channels $\bar B^0\to \pi^-\rho'^+$ and
$B^-\to \pi^-\rho'^0$ are lower as ones obtained by Ref. \cite{lipkin02}, although our value in the same 
order of magnitude that the only experimental branching ratio measured 
$B^-\to \pi^-\rho'^0$ \cite{babar05, babar09}.

\begin{acknowledgments}
The author acknowledge financial support from Promep, M\'exico. I thank to O. Navarro from Instituto 
de Investigaciones en Materiales, UNAM, M\'exico, for reading the manuscrip and his valuable 
suggestions.
\end{acknowledgments}

\begin{appendix}

\section{Matrix elements for $B\to P\pi'$ decays}

\ba {\cal M}(\bar B^0 \to \pi^-  \pi^{\prime+}) &=& -i{G_F\over \sqrt{2}}
f_{\pi^\prime} F_0^{B\to \pi} (m^2_{\pi^\prime}) (m_B^2-m^2_\pi)\nn\\
&&\left\{V_{ub}V^*_{ud}  a_1 -V_{tb}V^*_{td} \left[a_4+a_{10}+2(a_6+a_8)
{m^2_{\pi\prime}\over(m_b-m_u)(m_d+m_u)}\right]\right\}
\ea

\ba {\cal M}(\bar B^0 \to \pi^0 \pi^{\prime 0}) &=& i{G_F\over 2\sqrt{2}}
f_\pi F_0^{B\to \pi^\prime}(m_\pi^2) (m_B^2-m^2_{\pi^\prime})\nn\\
&& \left\{ V_{ub}V_{ud}^* a_2-V_{tb}V_{td}^* \left[-a_4+{1\over 2}a_{10}-{3\over 2}(a_7-a_9)
 -(2a_6-a_8){m^2_\pi \over(m_b-m_d)(m_d+m_d)} \right] \right\}\nn\\
&&+i{G_F\over \sqrt{2}} f_{\pi^\prime} F_0^{B\to \pi}(m^2_{\pi^\prime})(m_B^2-m^2_{\pi})\nn\\
&& \left\{V_{ub}V_{ud}^* a_2-V_{tb}V_{td}^* \left[-a_4+{1\over 2}a_{10}-{3\over 2}(a_7-a_9)
 -(2a_6-a_8){m^2_{\pi^\prime} \over(m_b-m_d)(m_d+m_d)}\right]\right\}
\ea

\ba {\cal M}(B^- \to \pi^- \pi^{\prime 0}) &=& -i{G_F\over 2} 
f_\pi F_0^{B\to \pi^\prime}(m_\pi^2) (m_B^2-m^2_{\pi^\prime})
\left\{V_{ub}V_{ud}^* a_1\right\}\nn\\
&&-i{G_F\over \sqrt{2}} f_{\pi^\prime} F_0^{B\to \pi}(m^2_{\pi^\prime})(m_B^2-m^2_\pi)\nn\\
&&\left\{V_{ub}V_{ud}^* a_2-V_{tb}V_{td}^*{3\over 2}
\left[a_9+a_{10}-a_7+2a_8{m^2_{\pi^\prime}\over(m_b-m_u)(m_d+m_u)}\right]\right\}
\ea

\ba {\cal M}( \bar B^0\to \etap \pi^{\prime 0}) &=& -i{G_F\over 2}
f_{\pi^\prime}F_0^{B\to \etap}(m^2_{\pi^\prime})(m^2_B-m^2_\etap)\nn\\
&&\left \{V_{ub}V_{ud}^* a_2 -V_{tb}V_{td}^*
\left[-a_4+{1\over 2}a_{10}-(2a_6-a_8){m^2_{\pi^\prime} \over(m_b-m_d)(m_d+m_d)} 
+\frac{3}{2}(a_9-a_7)\right]\right \} \nn\\
&+& i{G_F\over 2}f^u_\etap F_0^{B\to \pi}(m^2_\etap)(m^2_B-m^2_{\pi^\prime}) 
\left\{V_{ub}V_{ud}^* a_2+V_{cb}V_{cd}^*  a_2 {f^c_\etap\over f^u_\etap}
 \right .\nn \\
&&-V_{tb}V_{td}^*\left[a_4+2(a_3-a_5)+{1\over 2}(a_9-a_{7}-a_{10} )+(2a_6-a_8)
{m^2_\etap\over (m_b-m_d)(m_s+m_s)}
\left({f^s_\etap\over f^u_\etap}-1\right)r_\etap
\right. \nn\\
&&\left.\left.+(a_3-a_5+a_9-a_7){f^c_\etap\over f^u_\etap}+
\left(a_3-a_5 +{1\over 2}(a_7-a_9) \right)
{f^s_\etap\over f^u_\etap}\right]\right\}
\ea

\ba {\cal M}(B^- \to \etap \pi^{\prime -})  &=& i\frac{G_F}{\sqrt{2}}
f_{\pi^\prime} F_0^{B\to \etap}(m^2_{\pi^\prime})(m^2_B-m^2_{\etap})\nn\\
&&\left \{V_{ub}V_{ud}^* a_1 -V_{tb}V_{td}^*\left[a_4+a_{10} +2(a_6+a_8)
{m^2_{\pi^\prime} \over (m_b-m_u)(m_d+m_u)}\right]\right \}\nn\\
&& i\frac{G_F}{\sqrt{2}} f^u_\etap F_0^{B\to \pi^\prime}(m^2_{\etap}) 
(m^2_B-m^2_{\pi^\prime}) \left\{V_{ub}V_{ud}^* a_2 +V_{cb}V_{cd}^* 
a_2 {f^c_\etap \over f^u_\etap}\right .\nn \\
&& -V_{tb}V_{td}^*\left[a_4+2(a_3-a_5)+{1\over 2}(a_9-a_7-a_{10})+(2a_6-a_8)
{m^2_\etap \over(m_b-m_d)(m_s+m_s)}
\left({f^s_\etap \over f^u_\etap}-1\right)r_\etap \right. \nn \\
&&\left.\left.+(a_3-a_5+a_7-a_9){f^c_\etap\over f^u_\etap}
+\left(a_3-a_5 +{1\over 2}(a_7-a_9)\right ){f^s_\etap\over f^u_\etap}\right]\right \}
\ea

\ba {\cal M}(\bar B^0\to K^- \pi^{\prime +}) &=& -i\frac{G_F}{\sqrt{2}}
f_K F_0^{B\to \pi^\prime}(m^2_K)(m^2_B-m^2_{\pi^\prime})\nn\\
&&\left\{V_{ub}V_{us}^* a_1 -V_{tb}V_{ts}^*
\left[a_4+a_{10} +2(a_6+a_8){m^2_{K^-}\over (m_b-m_u)(m_u+m_s)}\right]\right\} 
\ea

\ba {\cal M}(\bar B^0 \to \bar K^0 \pi^{\prime 0}) &=& - i\frac{G_F}{{2}}
f_K F_0^{B\to \pi^\prime}(m^2_K)(m^2_B-m^2_{\pi^\prime})
 V_{tb}V_{ts}^* \left\{a_4-{1\over 2}a_{10} +(2a_6-a_8)
{m^2_{K^0}\over (m_b-m_d)(m_d+m_s)}\right\}\nn\\
&-& i\frac{G_F}{{2}} f_{\pi^\prime} F_0^{B\to K}(m^2_{\pi^\prime})(m^2_B-m^2_K) 
\left\{V_{ub}V_{us}^* a_2 -V_{tb}V_{ts}^* {3\over 2}(a_9-a_{7})\right \}
\ea

\ba {\cal M}(B^-\to K^- \pi^{\prime 0}) &=& -i{G_F\over 2}
f_K F_0^{B\to \pi^\prime}(m^2_K)(m^2_B-m^2_{\pi^\prime})
\left \{V_{ub}V_{us}^*a_1 - V_{tb}V_{ts}^*\left[a_4+a_{10} +2(a_6+a_8)
{m^2_{K^-}\over (m_b-m_u)(m_u+m_s)}\right]\right \}\nn\\
&-& i{G_F\over 2} f_{\pi^\prime} F_0^{B\to K}(m^2_{\pi^\prime})(m^2_B-m^2_K) 
\left\{V_{ub}V_{us}^* a_2 -V_{tb}V_{ts}^* {3\over 2}(a_9-a_7)\right \}
\ea

\ba {\cal M}(B^-\to \bar K^0 \pi^{\prime -}) &=& -i{G_F\over \sqrt{2}}
f_K F_0^{B\to \pi^\prime}(m^2_K)(m^2_B-m^2_{\pi^\prime})\nn \\
&& V_{tb}V_{ts}^* \left\{a_4-{1\over 2}a_{10}+(2a_6-a_8){m^2_{K^0}\over 
(m_b-m_d)(m_d+m_s)}\right\}
\ea

\section{Matrix elements for $B\to V\pi'$ decays}

\ba {\cal M}(\bar B^0\to \rho^-  \pi^{\prime+}) = \sqrt{2} G_F 
f_\rho F_1^{B\to \pi^\prime}(m^2_\rho) m_\rho (\epsilon \cdot p_{\pi^\prime} )
\left \{V_{ub}V_{ud}^* a_1  -V_{tb}V_{td}^*[a_4+a_{10}]\right \}
\ea

\ba {\cal M}(\bar B^0 \to \rho^+  \pi^{\prime-})  &=& \sqrt{2}G_F 
f_{\pi^\prime} A_0^{B\to \rho }(m^2_{\pi^\prime})
m_\rho (\epsilon \cdot p_{\pi^\prime} )\nn\\
&&\left\{V_{ub}V_{ud}^* a_1- V_{tb}V_{td}^* \left[a_4+a_{10}-2(a_6+a_8)
{m^2_{\pi^\prime}\over(m_b+m_u)(m_u+m_d)}\right]\right\}
\ea
 
\ba {\cal M}(\bar B^0\to \rho^0 \pi^{\prime 0}) &=& -{G_F\over \sqrt{2}} 
m_\rho (\epsilon \cdot p_{\pi^\prime}) \left( 
f_\rho F_1^{B\to \pi^\prime} (m^2_\rho)\left \{ V_{ub}V_{ud}^*a_2
+V_{tb}V_{td}^*\left[a_4-{1\over 2}a_{10}-{3\over 2}(a_7+a_9)\right]\right \}\right . \\
&+& \left . f_{\pi^\prime} A_0^{B\to \rho} (m^2_{\pi^\prime})
\left\{ V_{ub}V_{ud}^* a_2 +V_{tb}V_{td}^* \left[a_4-{1\over 2}a_{10}-(2a_6- a_8)
{m^2_{\pi^{\prime}}\over (m_b+m_d)(m_d+m_d)}+{3\over 2}(a_7-a_9) \right] \right\} \right)\nn
\ea

\ba {\cal M}(B^-\to \rho^0 \pi^{\prime-}) &=& G_F m_\rho 
(\epsilon \cdot p_{\pi^\prime}) \left (f_{\pi^\prime} A_0^{B\to \rho}
(m^2_{\pi^\prime}) \left \{ V_{ub}V_{ud}^* a_1 
-V_{tb}V_{td}^* \left[a_4+a_{10}-2(a_6+a_8){m^2_{\pi^\prime}\over(m_b+m_u)(m_u+m_d)}\right]
\right \}\right . \nn\\
&+& \left . f_\rho F_1^{B\to \pi^\prime }(m^2_\rho)
\left \{ V_{ub}V_{ud}^*a_2 -V_{tb}V_{td}^*
\left[-a_4+{1\over 2}a_{10}+{3\over 2}(a_7+a_9)\right]\right \}\right)
\ea

\ba {\cal M}( B^-\to \rho^- \pi^{\prime 0}) &=& G_F m_\rho 
(\epsilon \cdot p_{\pi^\prime}) \left(
f_\rho F_1^{B\to \pi^\prime } (m^2_\rho)
\left \{ V_{ub}V_{ud}^*  a_1 -V_{tb}V_{td}^* (a_4+a_{10})\right \}\right . \\
&+& \left . f_{\pi^\prime}A_0^{B\to \rho} (m^2_{\pi^\prime})
\left \{V_{ub}V_{ud}^* a_2 -V_{tb}V_{td}^* \left[-a_4+{1\over 2}a_{10}
-(2a_6-a_8){m^2_{\pi^{\prime}}\over (m_b+m_d)(m_d+m_d)}
+\frac{3}{2}(a_9-a_7)\right]\right \}  \right)\nn
\ea

\ba {\cal M}( \bar B^0 \to \omega \pi^{\prime 0})  &=& {G_F\over \sqrt{2}}
m_\omega(\epsilon \cdot p_{\pi^\prime})\left ( 
-f_\omega F_1^{B\to \pi^\prime } (m^2_\omega)\left \{ V_{ub}V_{ud}^*a_2
-V_{tb}V_{td}^*\left[a_4+2(a_3+a_5)+{1\over 2}(a_7+a_9-a_{10})\right] \right \}\right .\\
&+& \left . f_{\pi^\prime}A_0^{B\to \omega}(m^2_{\pi^\prime})
\left \{ V_{ub}V_{ud}^* a_2-V_{tb}V_{td}^*\left[-a_4+{1\over 2}a_{10}-(2a_6-a_8)
{m^2_{\pi^\prime}\over(m_b+m_u)(m_u+m_d)}+\frac{3}{2} (a_9-a_7)\right]\right \}\right)\nn
\ea

\ba {\cal M}( B^- \to \omega \pi^{\prime-}) &=& G_F m_\omega 
(\epsilon \cdot p_{\pi^\prime} ) \left( f_{\pi^\prime} A_0^{B\to \omega}
(m^2_{\pi^\prime})\left \{ V_{ub}V_{ud}^* a_1-V_{tb}V_{td}^*\left[a_4+a_{10}-2(a_6+a_8)
{m^2_{\pi^\prime}\over(m_b+m_u)(m_u+m_d)}\right]
\right \}\right. \nn\\
&+& f_\omega F_1^{B\to \pi^\prime}(m^2_\omega) \left \{ V_{ub}V_{ud}^*a_2 \right .
\left .\left .-V_{tb}V_{td}^*\left[a_4+2(a_3+a_5)+{1\over 2}(a_7+a_9-a_{10})
\right] \right \} \right)
\ea

\ba {\cal M}(\bar B^0\to K^{*-} \pi^{\prime+}) &=& \sqrt{2}G_F f_{K^*} 
F_1^{B\to \pi^\prime}(m^2_{K^*}) m_{K^*} (\epsilon \cdot p_{\pi^\prime}) \left 
\{V_{ub}V_{us}^* a_1 - V_{tb}V_{ts}^*[a_4+a_{10}]\right \} 
\ea

\ba {\cal M}(\bar B^0\to \bar K^{*0} \pi^{\prime 0})  &=&  G_F m_{K^{*0}} 
(\epsilon \cdot p_{\pi^\prime}) \left( f_{\pi^\prime} A_0^{B\to K^*}(m^2_{\pi^\prime})\left
\{V_{ub}V_{us}^* a_2 - V_{tb}V_{ts}^*{3\over 2}(a_9-a_7)\right \}\right .\nn\\
&+& \left . f_{K^*} F_1^{B\to \pi^\prime}(m^2_{K^{*0}})
V_{tb}V_{ts}^*  \left\{a_4-{1\over 2}a_{10}\right\} \right) 
\ea

\ba {\cal M}(B^-\to K^{*-} \pi^{\prime 0})  &=&  G_F m_{K^*} 
(\epsilon \cdot p_{\pi^\prime}) \left( f_{\pi^\prime}A_0^{B\to K^*}(m^2_{\pi^\prime})
\left \{V_{ub}V_{us}^* a_2 - V_{tb}V_{ts}^* {3\over 2}(a_9-a_7)\right \}\right .\nn\\
&+&\left . f_{K^*} F_1^{B\to \pi^\prime} (m^2_{K^*})\left \{
V_{ub}V_{us}^* a_1 -V_{tb}V_{ts}^* (a_4+a_{10}) \right \} \right)
\ea

\ba {\cal M}(B^-\to \bar K^{*0} \pi^{\prime-}) &=& -\sqrt{2} G_F f_{K^*} 
F_1^{B\to \pi^\prime} (m^2_{K^{*}}) m_{K^*} (\epsilon \cdot p_{\pi^\prime}) 
V_{tb}V_{ts}^*  \left\{a_4-\frac{1}{2}a_{10} \right\}
\ea

\ba {\cal M}(\bar B^0 \to \phi \pi^{\prime 0}) = -G_F f_\phi 
F_1^{B\to \pi^\prime}(m^2_\phi) m_\phi (\epsilon \cdot p_{\pi^\prime}) 
V_{tb}V_{td}^*  \left\{a_3+a_5-\frac{1}{2}(a_7+a_9)\right\} 
\ea

\ba {\cal M}(B^-\to \phi \pi^{\prime -}) &=& -\sqrt{2} 
{\cal M}(\bar B^0\to \phi \pi^{\prime 0}) 
\ea

\section{Matrix elements for $B\to P\rho'$ decays}

\ba {\cal M}(\bar B^0 \to \etap \rho^{\prime 0}) &=& G_F m_{\rho^\prime} 
(\epsilon \cdot p_{\etap})\left( f_{\rho^\prime} F_1^{B\to \etap}(m^2_{\rho^\prime}) 
\left \{V_{ub}V_{ud}^* a_2  \right.\right.\left.-V_{tb}V_{td}^* 
\left[-a_4+{1\over 2}a_{10}+{3\over 2}(a_7+a_9)\right]\right\}\nn\\
&+& f_{\etap}^u A_0^{B\to \rho^\prime}(m^2_\etap) \left\{V_{ub}V_{ud}^*  a_2 
+V_{cb}V_{cd}^*  a_2 \frac{f_{\etap}^c}{f_{\etap}^u} \right .\nn \\
&&-V_{tb}V_{td}^*\left[a_4+2(a_3-a_5)+{1\over 2}(a_9-a_7-a_{10})-(2a_6-a_8)
{m^2_{\etap}\over (m_b+m_d)(m_s+m_s)}
\left({f^s_\etap\over f^u_\etap}-1\right)r_\etap \right.\nn\\
&&~~~~\left.\left.\left.+(a_3-a_5-a_7+a_9)
{f^c_\etap\over f^u_\etap}+\left(a_3-a_5+{1\over 2}(a_7-a_9)\right)
{f^s_\etap\over f^u_\etap}\right] \right \}\right ) 
\ea

\ba {\cal M}(B^-\to \etap \rho^{\prime-}) &=& \sqrt{2}G_F m_{\rho^\prime} 
(\epsilon \cdot p_{\etap})\left(f_{\rho^\prime} F_1^{B\to \etap}(m^2_{\rho^\prime}) 
\left \{V_{ub}V_{ud}^* a_1 -V_{tb}V_{td}^*[a_4+a_{10} ]\right\}\right. \nn\\
&+& f^u_\etap A_0^{B\to \rho^\prime}(m^2_\etap) \left\{V_{ub}V_{ud}^* a_2 
+V_{cb}V_{cd}^* a_2 {f^c_\etap\over f^u_\etap} \right . \nn \\
&&-V_{tb}V_{td}^* \left[a_4+2(a_3-a_5)+{1\over 2}(a_9-a_7-a_{10})-(2a_6-a_8)
{m^2_\etap\over (m_b+m_d)(m_s+m_s)}
\left({f^u_\etap\over f^s_\etap}-1\right)r_\etap \right. \nn \\
&& \left.\left.\left.+(a_3-a_5-a_7+a_9){f^c_\etap\over f^u_\etap}+
\left(a_3-a_5-{1\over 2}(a_9-a_7) \right){f^s_\etap\over f^u_\etap}
\right] \right \}\right)
\ea

\ba {\cal M}(\bar B^0\to K^- \rho^{\prime +}) &=& \sqrt{2}G_F f_K 
A_0^{B\to \rho^\prime}(m^2_K) m_{\rho^\prime}(\epsilon \cdot p_K)\nn\\
&&\left \{V_{ub}V_{us}^* a_1-V_{tb}V_{ts}^* \left[a_4+a_{10}-2(a_6+a_8)
{m^2_{K^-}\over (m_b+m_u)(m_u+m_s)}\right]\right \}
\ea

\ba {\cal M}(\bar B^0\to \bar K^0 \rho^{\prime 0}) &=& G_F m_{\rho^\prime} 
(\epsilon \cdot p_K)\left( f_K A_0^{B\to \rho^\prime} (m^2_{K^0})
V_{tb}V_{ts}^* \left[a_4-{1\over 2}a_{10}-(2a_6-a_8)
{m^2_{K^0}\over (m_b+m_d)(m_d+m_s)}\right]\right.\nn\\
 &&+ \left.  f_{\rho^\prime} F_1^{B\to K}(m^2_{\rho^\prime})\left \{V_{ub}V_{us}^* a_2 
-V_{tb}V_{ts}^*{3\over 2}(a_7+a_9) \right \}\right) 
\ea

\ba {\cal M}(B^-\to K^- \rho^{\prime 0}) &=& G_F m_{\rho^\prime}
(\epsilon \cdot p_K) \left( f_K A_0^{B\to \rho^\prime} (m^2_K)
\left \{ V_{ub}V_{us}^*  a_1 -V_{tb}V_{ts}^* \left[a_4+a_{10}-2(a_6+a_8)
{m^2_{K^-}\over (m_b+m_u)(m_u+m_s)} \right]\right \}\right .\nn\\
&+&\left . f_{\rho^\prime} F_1^{B\to K}(m^2_{\rho^\prime})
\left \{ V_{ub}V_{us}^* a_2 - V_{tb}V_{ts}^*{3\over 2}(a_7+a_9) 
\right \} \right)
\ea

\ba {\cal M}(B^-\to \bar K^0 \rho^{\prime -}) &=& -\sqrt{2} G_F 
f_K A_0^{B\to \rho^\prime}(m^2_{K^0})m_{\rho^\prime}
(\epsilon \cdot p_K)\nn\\ 
&&V_{tb}V_{ts}^* \left\{a_4-{1\over 2}a_{10}-(2a_6-a_8)
{m^2_{K^0}\over (m_b+m_d)(m_d+m_s)}\right\}
\ea

\section{Matrix elements for $B\to V\rho'$ decays}

\ba {\cal M}(\bar B^0 \to \rho^-  \rho^{\prime+}) &=& 
X^{(\bar B^0 \rho^{\prime+}, \rho^-)}
\left\{V_{ub}V_{ud}^* a_1 - V_{tb}V_{td}^* [a_4+a_{10}]\right\}
\ea

\ba {\cal M}(\bar B^0 \to \rho^0 \rho^{\prime 0}) &=& 
\left[X^{(\bar B^0 \rho^0, \rho^{\prime 0})}+X^{(\bar B^0 \rho^{\prime 0}, \rho^0)}\right]
\left\{V_{ub}V_{ud}^* a_2 + V_{tb}V_{td}^*
\left[a_4-{1\over 2}a_{10}-{3\over 2}(a_7+a_9)\right]\right\}
\ea

\ba {\cal M}(B^-\to \rho^- \rho^{\prime 0}) &=&
X^{(B^-  \rho^-, \rho^{\prime 0})}\left\{V_{ub}V_{ud}^* a_2\right\} +
X^{(B^- \rho^{\prime 0}, \rho^-)}
\left\{V_{ub}V_{ud}^* a_1 -V_{tb}V_{td}^*
{3\over 2}(a_7+a_9+a_{10})\right \}
\ea

\ba {\cal M}(\bar B^0\to \omega \rho^{\prime 0}) &=&
X^{(\bar B^0 \rho^{\prime 0}, \omega)}
\left\{V_{ub}V_{ud}^* a_2 -V_{tb}V_{td}^* 
\left[a_4+2(a_3+a_5)+{1\over 2}(a_7+a_9-a_{10})\right]\right\}\nn\\ 
&+& X^{(\bar B^0 \omega, \rho^{\prime 0})}
\left\{V_{ub}V_{ud}^*  a_2 -V_{tb}V_{td}^* 
\left[-a_4+{1\over 2}a_{10}+{3\over 2}(a_7+a_9)\right]\right\}
\ea

\ba {\cal M}(B^-\to \omega \rho^{\prime -}) &=& 
X^{(B^- \omega, \rho^{\prime -})}
\left\{V_{ub}V_{ud}^* a_1 -V_{tb}V_{td}^* (a_4+a_{10}) \right \}\nn\\ 
&+&X^{(B^- \rho^{\prime -}, \omega)}
\left\{V_{ub}V_{ud}^*  a_2 -V_{tb}V_{td}^* \left[a_4+2(a_3+a_5)+
{1\over 2}(a_7+a_9-a_{10})\right] \right \}
\ea

\ba {\cal M}(\bar B^0 \to K^{*-} \rho^{\prime +}) &=& 
X^{(\bar B^0 \rho^{\prime +}, K^{*-})}
\left\{V_{ub}V_{us}^* a_1 -V_{tb}V_{ts}^* (a_4+a_{10})\right \}
\ea

\ba {\cal M}(\bar B^0\to \bar K^{*0} \rho^{\prime 0}) & =& 
X^{(\bar B^0 \bar K^{*0}, \rho^{\prime 0})}
\left\{V_{ub}V_{us}^* a_2-V_{tb}V_{ts}^* {3\over 2}(a_7+a_9)\right \} \nn\\
&+& X^{(\bar B^0 \rho^{\prime 0}, \bar K^{*0})}
\left\{-V_{tb}V_{ts}^*  \left[a_4-{1\over 2}a_{10}\right]\right\}
\ea

\ba {\cal M}(B^-\to K^{*-} \rho^{\prime 0}) &=&
X^{(B^- K^{*-}, \rho^{\prime 0})}
\left\{V_{ub}V_{us}^* a_2 -V_{tb}V_{ts}^* {3\over 2}(a_7+a_9)\right \}\\
&+&X^{(B^- \rho^{\prime 0}, K^{*-})}
\left\{V_{ub}V_{us}^* a_1 -V_{tb}V_{ts}^* (a_4+a_{10})\right \}\nn
\ea

\ba {\cal M}( B^-\to \bar K^{*0}\rho^{\prime-}) &=& 
-X^{(B^- \rho^{\prime -}, \bar K^{*0})}
V_{tb}V_{ts}^* \left \{a_4-{1\over 2}a_{10} \right \}
\ea

\ba {\cal M}(\bar B^0\to \phi \rho^{\prime 0})&=& 
X^{(\bar B^0 \rho^{\prime 0}, \phi)}
\left\{V_{tb}V_{td}^*\left[a_3+a_5-{1\over 2}(a_7+a_9)\right]\right\}
\ea

\ba {\cal M}( B^-\to \phi \rho^{\prime -}) &=&-\sqrt{2} {\cal M}(\bar B^0\to \phi \rho^{\prime 0})
\ea

\end{appendix}

\end{document}